\documentclass[aps,preprint,showpacs,preprintnumbers,amsmath,amssymb]{revtex4}
\usepackage[dvips]{graphicx}
\usepackage{float}
\begin{document}

\title{Interaction between Maxwell field and charged scalar field in de Sitter universe}
\author{Cosmin Crucean \thanks{E-mail:~~crucean@physics.uvt.ro}
and Mihaela-Andreea B\u aloi \thanks{E-mail:~~mihaela.baloi88@e-uvt.ro}\\
{\small \it West University of Timi\c soara,}\\
{\small \it V. Parvan Ave. 4 RO-300223 Timi\c soara,  Romania}}

\begin{abstract}
We study the theory of interaction between charged scalar field and Maxwell field in de Sitter background. Solving the equation of interacting fields we define the in-out fields as asymptotic free fields and construct the reduction formalism for scalar field. Then we derive the perturbation expansion of the scattering operator. The first order transition amplitudes corresponding to particle production from de Sitter vacuum and pair production in an external field are analysed. We show that all these effects are important only in strong gravitational fields and vanish in the flat limit.
\end{abstract}


\maketitle
\section{\label{sec:1}Introduction}

In de Sitter geometry the problem of particle production was considered by many authors. These studies show that the space expansion generate particle production using various methods \cite{co3,Cs,Bbl,rb,WKB,wkb1,wkb2,sr,B16,B17,B18,B19,B20,B21,B22,B23,Ts}. However a detailed study of interaction between scalar field and Maxwell field minimally coupled with gravitation using perturbative methods and exact calculations received little attention. We focus our attention on this subject because the last astronomical observations shows that the expansion of the Universe is accelerating, thus increasing the interest on de Sitter model. The study of interactions between bosons and scalar fields on a curved background will help us to understand how the scalar electrodynamics from flat space will be modified when a gravitational field is present. Important results such as normalised solutions was obtained for Dirac field \cite{Br},\cite{co1} and scalar field \cite{co2},\cite{sr},\cite{B14,B15} on de Sitter geometry. In the case of the electromagnetic field the situation is simple if one recalls the conformal invariance between de Sitter and Minkowski metrics. This means that the theory of free electromagnetic field is the same as in Minkowski geometry if we use the metric with conformal time \cite{Max}. In the present paper our we will study on the interaction between charged scalar field and electromagnetic field.

We want to study the effect of electromagnetic interaction upon particle production in de Sitter expanding universe. The intention of this paper is to develop the theory of the scalar QED on de Sitter spacetime and to use this formalism for studying the following processes: $vacuum \rightarrow \varphi + \varphi^{*}$ in an external Coulomb field and $vacuum \rightarrow \varphi + \varphi^{*} + \gamma$. Here $\varphi$ represents the charged scalar particle, $\varphi^{*}$ represents the charged scalar antiparticle and $\gamma$ the photon. The annihilation of scalar pair into vacuum in the presence of Coulomb field and annihilation of the triplet scalar pair and a photon into de Sitter vacuum represents the time reversed processes. Our results prove that these probabilities are nonvanishing only in strong gravitational fields. Our exact calculations allows us to study the behaviour of our amplitudes/probabilities for large/small values of the expansion factor, making in this way the connection with the scalar QED from Minkowski space. We must specify that in our calculations we will use the solutions of Klein-Gordon and Maxwell equations which have a defined momentum. These reactions are forbidden in flat space theory because of the energy-momentum conservation laws. In de Sitter case the energy is no longer conserved and we must expect to obtain a nonvanishing probability for these processes.

We start in the second section with a review of the results concerning the free scalar field theory on de Sitter space. In the third section we construct the theory of interaction between charged scalar field and electromagnetic field on de Sitter space and then our attention will be focused on developing the reduction formalism for scalar field. In this section is also studied the perturbation theory which help us to obtain the definition of scattering amplitudes. The first order processes that generate particle production will be studied in sections four and five. Further we study the probabilities that generate particle production using a graphical method. Our conclusions are summarized in section six, and in the Appendix we give the main steps that help us to calculate the transition amplitudes.

\section{\label{sec:2}Free scalar field}

The exact solutions of the free Klein-Gordon equation on de Sitter space-time, was obtained in \cite{co2},\cite{sr}. The calculations was done using the de Sitter line element \cite{B1,B2}:
\begin{equation}\label{metr}
ds^{2}=dt^{2}-e^{2\omega t}d\vec{x}^{\,2}=\frac{1}{(\omega t_{c})^2}\,\left(dt^{2}_{c}- d\vec{x}\,^2\right),
\end{equation}
where $\omega$ is the expansion factor.
In the chart $\{t,\vec{x}\}$ with Cartesian coordinates the Klein-Gordon
equation reads \cite{co2}:
\begin{equation}
\left( \partial_t^2-e^{-2\omega t}\Delta +3\omega
\partial_t+m^2\right)\phi(x)=0\,.\label{eq:KG1}
\end{equation}
It is known that the Klein-Gordon equation (\ref{eq:KG1}) on de Sitter space can be analytically solved
in terms of Bessel functions \cite{B14},\cite{sr}. In our study we will use the
normalized solutions of positive frequencies obtained in \cite{co2}:
\begin{equation}
f_{\vec{
p}}\,(x)=\frac{1}{2}\sqrt{\frac{\pi}{\omega}}\,\frac{e^{-3\omega
t/2}}{(2\pi)^{3/2}}\,e^{-\pi k/2}H^{(1)}_{ik}\left(\frac{p}{\omega}\,e^{-\omega t}\right) e^{i \vec{p}\cdot\vec{
x}}\,,\label{eq:sol}
\end{equation}
where $H^{1}_{\mu}(z)$ is the Hankel function of first kind, $p=|\vec{p}|$ and the notations are:
\begin{equation}
k=\sqrt{u^2-\textstyle{\frac{9}{4}}}\,, \quad u=\frac{m}{\omega}\,,
\label{eq:k}
\end{equation}
with $m>3\omega/2$. The fundamental solutions of negative
frequencies are $f_{\vec{p}}^*(x)$.
For $t \rightarrow -\infty$, the modes defined above behave as a positive/negative frequency modes with respect to the conformal time $t_{c}=-e^{-\omega t}/\omega$:
\begin{equation}
f_{\vec{p}}\,(x)\sim e^{-ipt_{c}},\,\,\,f_{\vec{p}}^*\,(x)\sim e^{ipt_{c}}
\end{equation}
This is the behavior in the infinite past for positive/negative frequency modes
which defines the Bunch-Davies vacuum.

The fundamental solutions of Klein-Gordon equation, satisfy the ortonormalizations relations \cite{co2}:
\begin{eqnarray}
i\int d^3x\, (-g)^{1/2}\, f_{\vec{p}}^*(x)
\stackrel{\leftrightarrow}{\partial_{t}} f_{\vec{p}^{\,\,\prime}}(x)&=&\nonumber\\
-i\int d^3x\, (-g)^{1/2}\, f_{\vec{p}}\,(x)
\stackrel{\leftrightarrow}{\partial_{t}} f^*_{\vec{p}^{\,\,\prime}}(x)
&=&\delta^3(\vec{p}-\vec{p}^{\,\,\prime})\,,\\
i\int d^3x\, (-g)^{1/2}\, f_{\vec{p}}\,(x)
\stackrel{\leftrightarrow}{\partial_{t}} f_{\vec{p}^{\,\,\prime}}(x)&=&0\,,
\end{eqnarray}
where the integration extends on an arbitrary hypersurface
$t=const$ and $(-g)^{1/2}=e^{3\omega t}$. 
The completeness condition satisfied by these solutions reads \cite{co2}:
\begin{equation}
i\int d^3p\,  f^*_{\vec{p}}(t,\vec{x}\,) \stackrel{\leftrightarrow}{\partial_{t}}
f_{\vec{p}}\,(t,\vec{x}^{\,\,\prime})=e^{-3\omega t}\delta^3(\vec{x}-\vec{x}^{\,\,\prime})\,.\label{eq:comp}
\end{equation}

The quantization can be done considering the plane wave in
momentum representation:
\begin{equation}
\phi(x)=\phi^{(+)}(x)+\phi^{(-)}(x)=\int d^3p \left[f_{\vec{p}}\,(x)a(\vec{p}\,)+f_{\vec{p}}^*(x)b^+(\vec{p}\,)\right],\label{eq:field1}
\end{equation}
where the particle $(a,a^{\dagger})$ and antiparticle ($b,b^{\dagger})$
operators satisfy the standard commutation relations in
momentum representation:
\begin{equation}
[a(\vec{p}\,), a^{\dagger}(\vec{p}^{\,\,\prime})]=[b(\vec{p}\,), b^{\dagger}(\vec{p}^{\,\,\prime})] = \delta^3 ({\vec{p}}-\vec{p}^{\,\,\prime})\,.\label{eq:com1}
\end{equation}
These operators can be calculated using the inversion formulas:
\begin{eqnarray}
a(\vec{p}\,)&=&i\int  d^{3}x(-g)^{1/2} f^*_{\vec{p}}(x)\stackrel{\leftrightarrow}{\partial_{t}}\phi(x) \nonumber\\
b(\vec{p}\,)&=&i\int  d^{3}x (-g)^{1/2} f^*_{\vec{p}}(x)\stackrel{\leftrightarrow}{\partial_{t}} \phi^{\dagger}(x)\label{eq:inv}.
\end{eqnarray}

All these operators act on the Fock space
supposed to have an unique vacuum state $|0\rangle$ accomplishing
\begin{equation}
a(\vec{p}\,)|0\rangle=b(\vec{p}\,)|0\rangle=0\,,\quad \langle0|a^{\dagger}(\vec{
p}\,)=\langle 0|  b^{\dagger}(\vec{p}\,)=0\,.
\end{equation}

In this way  the field $\phi$ is
correctly quantized according to the canonical rule \cite{co2}:
\begin{equation}
[ \phi(t,\vec{
x}\,),\partial_{t}\phi^{\dagger}(t,\vec{x}^{\,\,\prime})]=ie^{-3\omega t}\delta^3(\vec{x}-\vec{x}^{\,\,\prime})\,.
\end{equation}
In the quantum theory of fields it is important to study  the Green functions
related to the partial commutator functions (of positive or negative
frequencies) defined as:
\begin{equation}
D^{(\pm)}(x,x')= i[\phi^{(\pm)}(x),\phi^{(\pm) \dagger}(x')],
\end{equation}
and the total one, $D=D^{(+)}+D^{(-)}$. These functions are solutions of the Klein-Gordon
equation in both sets of variables and obey
$[D^{(\pm)}(x,x')]^*=D^{(\mp)}(x,x')$ . Then the partial commutator functions will have the expressions:
\begin{eqnarray}
D^{(+)}(x,x')=i\int d^3 p \, f_{\vec{p}}\,(x)f^*_{\vec{p}}(x') ,\nonumber\\
D^{(-)}(x,x')=-i\int d^3 p \, f_{\vec{p}}^*(x)f_{\vec{p}}\,(x').
\end{eqnarray}
These functions help us to introduce the principal Green functions.  In
general, $G(x,x')=G(t,t',\vec{x}-\vec{x}^{\,\,\prime})$ is a Green function of the Klein-Gordon
equation if satisfies \cite{co2}:
\begin{equation}
\left( \partial_t^2-e^{-2\omega t}\Delta_x +3\omega
\partial_t+m^2\right)G(x,x')=e^{-3\omega t}\delta^4(x-x')\,.
\label{eq:KGG}
\end{equation}
The properties of the commutator functions allow us to construct the Green
function just as in the scalar theory on Minkowski spacetime. We assume that
the retarded, $D_R$, and advanced, $D_A$, Green functions read:
\begin{eqnarray}
D_R(t,t',\vec{x}-\vec{x}^{\,\,\prime})&=& \theta(t-t')D(t,t',\vec{x}-\vec{x}^{\,\,\prime})\,,\\
D_A(t,t',\vec{x}-\vec{x}^{\,\,\prime})&=& -\,\theta(t'-t)D(t,t',\vec{x}-\vec{x}^{\,\,\prime})\,,
\label{eq:RA}
\end{eqnarray}
while the Feynman propagator,
\begin{eqnarray}
D_F(t,t',\vec{x}-\vec{x}^{\,\,\prime})&=& i\langle 0|T[\phi(x)\phi^{\dagger}(x')]\,|0\rangle\nonumber\\
&=& \theta (t-t') D^{(+)}(t,t',\vec{x}-\vec{x}^{\,\,\prime})-\theta(t'-t)D^{(-)}(t,t',\vec{x}-\vec{x}^{\,\,\prime})\,,
\end{eqnarray}
is defined as a causal Green function. Thus we have now all the elements that we need for begin our study related to the reduction formulas for scalar field on de Sitter space-time.

\section{Scalar QED}

The goal of the quantum field theory is to describe the dynamics of interacting particles. We are interested to obtain the amplitudes for single particles to propagate in a curved space-time and the transition amplitudes for interacting particles between different initial and final states.
Our main attention in this paper will be focused on interaction between charged massive scalar field and
electromagnetic field on de Sitter space-time.
\subsection{Interacting fields}
In this work we use for our calculation the same formalism as in \cite{B3}-\cite{B8}, and we adopt the minimal coupling corresponding to the classical interaction of a point charge as the prescription for introducing electrodynamic couplings. Also we will note the interacting fields with $\phi(x)$ and $A_{\mu}(x)$. These fields will satisfy a system of equations that can be obtained from the following Lagrangian density:
\begin{eqnarray}
\mathcal{S}[\phi,A]&=&\int d^{4}x[\mathcal{L}_{SC}(\phi)+\mathcal{L}_{M}(A)+\mathcal{L}_{I}(\phi,A)]\nonumber\\
&&=\sqrt{-g}\biggl\{(\partial^{\mu}\phi^{\dagger}\partial_{\mu}\phi-m^2 \phi^{\dagger} \phi)-\frac{1}{4}F_{\mu\nu}F^{\mu\nu}\nonumber\\
&&-ie[(\partial_{\mu}\phi^{\dagger})\phi-(\partial_{\mu}\phi)\phi^{\dagger}]A^{\mu} +e^{2}\phi^{\dagger}\phi A_{\mu}A^{\mu}\biggl\},\label{eq:action}
\end{eqnarray}
where the density of lagrangian of interaction reads $\mathcal{L}_{I}=-ie\sqrt{-g}[(\partial_{\mu}\phi^{\dagger})\phi-(\partial_{\mu}\phi)\phi^{\dagger}]A^{\mu} +\sqrt{-g}e^{2}\phi^{\dagger}\phi A_{\mu}A^{\mu}$. This relation is obtained after we make the substitution $\partial_{\mu}\phi \rightarrow \partial_{\mu}\phi-ieA_{\mu}\phi$ in the lagrangian of free scalar field.
From action (\ref{eq:action}), we obtain the system of equations for interacting fields on a curved background as follows:
\begin{eqnarray}
&&\frac{1}{\sqrt{(-g)}}\partial_{\mu}\left(\sqrt{(-g)}F^{\mu\nu}\right)=-ie[(\partial^{\nu}\phi^{\dagger})\phi-(\partial^{\nu}\phi)\phi^{\dagger}]
+2e^{2}\phi^{\dagger} \phi A^{\nu},\nonumber\\
&&\frac{1}{\sqrt{-g}}\,\partial_{\mu}\left[\sqrt{-g}\,
g^{\mu\nu}\partial_{\nu}\phi-ie\sqrt{-g}\phi A^{\mu}\right]+m^2\phi=ie(\partial_{\mu}\phi)A^{\mu}+e^{2}\phi A_{\mu}A^{\mu}.
\label{eq:eqs}
\end{eqnarray}
The problem of finding exact solutions for the system (\ref{eq:eqs}), which contain coupled nonlinear equations for interacting fields proves to be as in the flat space case, a complicated task. It is immediate from here that for finding solutions of the coupled equations, we can try to follow the same approximation method as in flat space case \cite{B3}-\cite{B8}.
The system of coupled equations is replaced with one system of integral equations that contain information about initial conditions. For that we select the Green function $G(x,y)$ corresponding to one initial condition, which help us
to write the solution of the second equation from system (\ref{eq:eqs}) as follows:
\begin{eqnarray}
\phi(x)&=&\hat{\phi}(x)+e\int
d^{4}y\sqrt{-g}\,G(x,y)\biggl\{\frac{i}{\sqrt{-g}}\partial_{\mu}[\sqrt{-g}\phi(y) A^{\mu}(y)]\nonumber\\
&&+i[\partial_{\mu}\phi(y)]A^{\mu}(y)+e\phi(y) A_{\mu}(y)A^{\mu}(y)\biggl\},\label{eq:int}
\end{eqnarray}
where $\hat{\phi}(x)$  are free fields. One can verify that (\ref{eq:int}) is the exact
solution of the second equation from (\ref{eq:eqs}), applying $(E_{KG}+m^{2})$
to $\phi(x)$, with the observation that
$[E_{KG}+m^{2}]\hat{\phi}(x)=0$. We introduce the following convenable notation $E_{KG}=\frac{1}{\sqrt{-g}}\,\partial_{\mu}(\sqrt{-g}\,
g^{\mu\nu}\partial_{\nu})$. Using the fact that Green functions must satisfy equation (\ref{eq:KGG}), it is not hard to verify that the solution (\ref{eq:int}) is justified.

Since our main scope in this paper is to study transition amplitudes, we want to construct the states which give a simple description of the physical system at the time $t\rightarrow -\infty$. At this time the particles have not yet interacted with each other and propagate under the influence of their self-interactions only. This implies that for constructing a theory of interactions, we must seek for operators that create independent particle states with each particle propagating with its physical mass.
Let us start to construct these operators, using  (\ref{eq:int}) which offers us the possibility of constructing free fields, which are asymptotic equal (at $t\rightarrow\pm\infty$)
with solutions of system (\ref{eq:eqs}). As we know the retarded
Green functions $D_{R}(x,y)$ vanishes at $t\rightarrow-\infty$
while the advanced one $D_{A}(x,y)$ vanishes for
$t\rightarrow\infty$. Thus equation (\ref{eq:int}) can be expressed with
retarded and advanced functions as follows:
\begin{eqnarray}
\phi(x)&=&\hat{\phi}_{R/A}(x)+e\int
d^{4}y\sqrt{-g}D_{R/A}(x,y)\biggl\{\frac{i}{\sqrt{-g}}\partial_{\mu}[\sqrt{-g}\phi(y) A^{\mu}(y)]\nonumber\\
&&+i[\partial_{\mu}\phi(y)]A^{\mu}(y)+e\phi(y) A_{\mu}(y)A^{\mu}(y)\biggl\}\label{eq:phi}.
\end{eqnarray}
Now we can define the free fields, $\hat{\phi}_{R/A}(x)$ that satisfy:
\begin{eqnarray}
\lim_{t\rightarrow\mp\infty}(\phi(x)-\hat{\phi}_{R/A}(x))=0.
\end{eqnarray}
The free fields $\hat{\phi_{R}}$ and $\hat{\phi_{A}}$ have mass
$m$ and are equal at $t\pm\infty$ with exact solutions of the
coupled equations and represent the fields before and after the
interaction. It is known that the mass of
the interacting fields may differ from the mass of free fields. For obtaining the mass correction we will ad to the lagrangian of interaction a term of the form $-\delta m^{2}\phi^{\dagger} \phi$, which help us to rewrite the second equation from system (\ref{eq:eqs}):
\begin{equation}
\frac{1}{\sqrt{-g}}\,\partial_{\mu}\left[\sqrt{-g}\,
g^{\mu\nu}\partial_{\nu}\phi-ie\sqrt{-g}\phi A^{\mu}\right]+m^2\phi=ie(\partial_{\mu}\phi)A^{\mu}+e^{2}\phi A_{\mu}A^{\mu}-\delta m^{2}\phi,\label{eq:dm}
\end{equation}
where the current $\widetilde{j}=ie(\partial_{\mu}\phi)A^{\mu}+e^{2}\phi A_{\mu}A^{\mu}-\delta m^{2}\phi+\frac{ie}{\sqrt{-g}}\partial_{\mu}(\sqrt{-g}\phi A^{\mu})$ is now the source giving rise to the scalar fields.

Further we can follow the method from Minkowski case \cite{B3}-\cite{B8}, where the free fields $\hat{\phi_{R}}$ and
$\hat{\phi_{A}}$ are defined up to a normalization constant noted
with $\sqrt{z}$. This allows us to define the $\emph{in}/\emph{out}$ fields in order to construct the states of single particles, as follows:
\begin{eqnarray}
\sqrt{z}\phi_{in/out}(x)&=&\phi(x)-e\int
d^{4}y\sqrt{-g}D_{R/A}(x,y)\biggl\{\frac{i}{\sqrt{-g}}\partial_{\mu}[\sqrt{-g}\phi(y) A^{\mu}(y)]\nonumber\\
&&+i[\partial_{\mu}\phi(y)]A^{\mu}(y)+e\phi(y) A_{\mu}(y)A^{\mu}(y)-\delta m^{2}\phi(y)\biggl\}.
\label{eq:inout1}
\end{eqnarray}
The $\emph{in}/\emph{out}$ free fields defined above satisfy Klein-Gordon
equation, are written with the help of creation and
annihilation operators and in addition satisfy the conditions:
\begin{equation}
\lim_{t\rightarrow\mp\infty}(\phi(x)-\sqrt{z}\phi_{in/out}(x))=0.
\end{equation}
It is useful to express these fields with the help of equation (\ref{eq:dm}), in the form:
\begin{equation}
\sqrt{z}\phi_{in/out}(x)=\phi(x)-\int
d^{4}y\sqrt{-g}D_{R/A}(x,y)[E_{KG}(y)+m^{2}]\phi(y).\label{eq:inout2}
\end{equation}
From the above normalization we can define the creation and annihilations operators as in Section II ,
\begin{eqnarray}
a(\vec{p}\,)_{in/out}&=&i\int  d^{3}x(-g)^{1/2} f^*_{\vec{p}}(x)\stackrel{\leftrightarrow}{\partial_{t}}\phi_{in/out}(x) \nonumber\\
b^{\dagger}(\vec{p}\,)_{in/out}&=&i\int  d^{3}x (-g)^{1/2} f_{\vec{p}}\,(x)\stackrel{\leftrightarrow}{\partial_{t}} \phi_{in/out}(x).\label{eq:ab}
\end{eqnarray}
The creation and annihilation operators defined above satisfy the
commutation relations (\ref{eq:com1}) and from that it follows that all
the properties of free fields will be preserved.

Before starting our calculations we make a few remarks about the
scattering operator. Denoting the vacuum state $|0\rangle$ (which is suppose to be unique), the
one particle/antiparticle states can be written:
\begin{eqnarray}
a^{\dagger}(\vec{p}\,)_{in/out}|0\rangle&=&|in/out,1(\vec{p}\,)\rangle\nonumber\\
b^{\dagger}(\vec{p}\,)_{in/out}|0\rangle&=&|in/out,\widetilde{1}(\vec{p}\,)\rangle.
\end{eqnarray}
If we consider two states $|in,\alpha\rangle$ and
$|out,\beta\rangle$, then the probability of
transition from state $\alpha$ to state $\beta$ is defined as the scalar
product of the two states: $\langle out,\beta|in,\alpha\rangle$.
These are just the elements of matrix for scattering operator $S_{\beta\alpha}$, which are of physical interest. This
operator assures the stability of the vacuum state and one
particle state, and in addition transform any $out$ field in the
equivalent $in$ field.

Our remaining task is to construct and study, general matrix
elements which describe the dynamical behavior of interacting
particles.  As in Minkowski theory \cite{B3}-\cite{B8}, we can construct
$n-$particle $in$ and $out$ states as in the free scalar field theory by
repeated application to the vacuum of
$a^{\dagger}(\vec{p}\,)_{in/out}$ and
$b^{\dagger}(\vec{p}\,)_{in/out}$.

\subsection{Reduction formalism}
The reduction formulas for scalar particles can be obtained as in Minkowski theory \cite{B3}-\cite{B8}. Let us consider the amplitude for the process in which particles
from $in$ states denoted with $\alpha$, together with a scalar particle $1(\vec{p}\,)$ pass in $out$ state, in which we have a scalar particle $1(\vec{p}^{\,\,\prime})$ and particles denoted by
$\beta$. The reduction from out state can be done using the following relations:
\begin{eqnarray}
a_{out}(\vec{p}^{\,\,\prime})-a_{in}(\vec{p}^{\,\,\prime})=i\int
d^{3}x e^{3\omega t}f^*_{\vec{p}^{\,\,\prime}}(x)\stackrel{\leftrightarrow}{\partial_{t}}
(\phi_{out}(x)-\phi_{in}(x)),\label{eq:aa}
\end{eqnarray}
\begin{equation}
\phi_{out}(x)-\phi_{in}(x)=\frac{1}{\sqrt{z}}\int
d^{4}y\sqrt{-g} D(x,y)[E_{KG}(y)+m^{2}]\phi(y),\label{eq:dif}
\end{equation}
where we noted $D(x,y)=D_{R}(x,y)-D_{A}(x,y)$.
The final result for reduction of scalar particle from $out$ state is:
\begin{eqnarray}
\langle out \beta,1(\vec{p}^{\,\,\prime})|in
\alpha,1(\vec{p}\,)\rangle&=&\delta^{3}(\vec{p}-\vec{p}^{\,\,\prime})\langle out
\beta|in \alpha\rangle\nonumber\\
&&+\frac{i}{\sqrt{z}}\int
\sqrt{-g}f^*_{\vec{p}^{\,\,\prime}}(y) [E_{KG}(y)+m^{2}]\nonumber\\
&&\times\langle out
\beta|\phi(y)|in
\alpha,1(\vec{p}\,)\rangle d^{4}y.\label{eq:fout}
\end{eqnarray}

This method can be also applied for the reduction of particles from $in$ state and the final result read:

\begin{eqnarray}
\langle out \beta,1(\vec{p}^{\,\,\prime})|in
\alpha,1(\vec{p}\,)\rangle&=&\delta^{3}(\vec{p}-\vec{p}^{\,\,\prime})\langle out
\beta|in \alpha\rangle\nonumber\\
&&+\frac{i}{\sqrt{z}}\int
\sqrt{-g}\langle out \beta,1(\vec{p}^{\,\,\prime})|\phi^{\dagger}(y)|in
\alpha,1(\vec{p}\,)\rangle\nonumber\\
&&\times  [E_{KG}(y)+m^{2}]f_{\vec{p}}\,(y)d^{4}y.
\end{eqnarray}

Let us resume the above results related to reduction method. We showed that
all particles, will be replaced by formulas that were obtained in the
reduction of one particle. When more particles are reduced,
in the matrix element appears the time ordered products of
corresponding field operators. Every particle will be replaced
after we apply the above reduction method with expressions which depend on the field operator
$\phi(x)$. As in the flat space case, we can associate a
Green function to any process of interaction that must be calculated using perturbation theory.

\subsection{Perturbation theory}
Expanding the transition amplitudes and matrix elements  in a power series in the interaction strength, we can obtain the rules for computing the terms in the expansion. The Green functions of the interacting fields cannot be calculated
exact and for that reason we must use perturbation methods. The
form of the amplitudes obtained from reduction formalism  allows
us to use perturbation calculus as in Minkowski theory \cite{B3}-\cite{B8}.

For calculating the Green functions it will be helpful to write them in terms of free fields as follows:
\begin{equation}
G(y_{1},y_{2},...,y_{n})=\langle0|T[\hat{\phi}(y_{1})\hat{\phi}(y_{2})...\hat{A}(y_{n})]|0\rangle=
\frac{\langle0|T[\hat{\phi}(y_{1})\hat{\phi}(y_{2})...\hat{A}(y_{n}),\mathbf{S}]|0\rangle}{\langle0|\mathbf{S}|0\rangle}.
\label{eq:G}
\end{equation}
Then the entire perturbation calculus is based on expansion of
the operator $\mathbf{S}$:
\begin{eqnarray}
\mathbf{S}=T e^{-i\int
\mathcal{L}_{I}(x)d^{4}x}=1+
\sum\limits_{n=1}^\infty
\frac{(-i)^{n}}{n!}\int T[\mathcal{L}_{I}(x_{1})...\mathcal{L}_{I}(x_{n})]d^{4}x_{1}...d^{4}x_{n},
\label{eq:S}
\end{eqnarray}
where the density lagrangian of interaction is (we neglect the second term in the density of lagrangian of interaction):
$\mathcal{L}_{I}(x)=-i e\sqrt{-g}
:[(\partial_{\mu}\phi^{\dagger})\phi-(\partial_{\mu}\phi)\phi^{\dagger}]A^{\mu}:$.
Each term from (\ref{eq:S}) corresponds to an
order from perturbation theory. Replacing (\ref{eq:S}) in the expression of
Green function (\ref{eq:G}) we obtain perturbation series which allows
us to calculate the transition amplitude in any order.

With this method, the definition of transition
amplitudes in any order of perturbation theory can be obtained. The Green functions from equation (\ref{eq:G}) can be expressed in terms of Feynmann propagators considering all the T contractions from Wick theorem.
For example in the first order of perturbation theory the transition amplitude reads:

\begin{equation}
\langle(\vec{p\,'})|\mathbf{S}_{1}|(\vec{p}\,)\,\rangle=-e
\int\sqrt{-g(x)}\left[f^*_{\vec{p}^{\,\,\prime}}(x)\stackrel{\leftrightarrow}{\partial_{\mu}}f_{\vec{p} }\,(x)\right]A^{\mu}(x)d^{4}x,
\label{eq:ampl2}
\end{equation}
where the bilateral derivative acts as: $g\stackrel{\leftrightarrow}{\partial}h=(\partial h)g-(\partial g)h$.
The above expression is just the scattering amplitude in the first order of the perturbation theory on de Sitter expanding universe. With the above method were established the
general rules of calculation for deducing the expression for the scattering amplitudes in any order of the perturbation theory. The results obtained in the last two sections will be applied to calculate the processes which generate particle production.

\section{Particle production in Coulomb field}
In this section we want to study the process $vacuum\rightarrow\varphi+\varphi^{*}$, in the presence of Coulomb field, using the scalar QED formalism developed in the previous sections. More precisely, we want to show that there are nonvanishing probabilities for scalar pair production in the case of expanding de Sitter universe.
It is well known that the Coulomb field could not generate particle production in scalar QED from Minkowski space as a perturbative phenomenon. Contrary to this, in the de Sitter case we must expect that this process to be allowed by the fact that energy is no longer conserved. The studies done before with fermions indeed show that there are nonvanishing probabilities for pair generation in external Coulomb field \cite{cd}. The ingredients of our calculations will be the solutions of Klein-Gordon equation in the momentum basis and the form of Coulomb field in de Sitter geometry. Using the conformal invariance we can establish the expression for Coulomb field in this geometry in the natural frame \cite{cd,Max}:
\begin{equation}
A^{0}(x)=\frac{Ze}{|\vec{x}|}\,e^{-2\omega
t},\,\,\,A^{i}=0.\label{eq:pot}
\end{equation}

The amplitude corresponding to scalar pair production in Coulomb field on de Sitter spacetime, reads:
\begin{equation}
\langle(\vec{p}\,)_{-},(\vec{p\,'})_{+}|\mathbf{S}_{1}|0\,\rangle=-e
\int\sqrt{-g(x)}\left[f^*_{\vec{p}^{\,\,\prime}}(x)\stackrel{\leftrightarrow}{\partial_{t}}f^*_{\vec{p} }\,(x)\right]A^{0}(x)d^{4}x,
\label{eq:ampl2}
\end{equation}
where the notations $(\vec{p}\,)_{-}$ stands for a scalar particle of momentum $\vec{p}$ and negative charge and $(\vec{p\,'})_{+}$ denotes the antiparticle with positive charge and momentum $\vec{p\,'}$.

\subsection{Probability of pair production}
In order to obtain the expression of the probability of pair production, we first need to calculate the amplitude of transition
(\ref{eq:ampl2}).
We use in our calculations the solutions of Klein-Gordon equation in momentum basis (\ref{eq:sol}). These solutions are written in terms of Hankel functions with imaginary index, provided that $m/\omega>3/2$. The spatial integral is the same as in the flat case and the result is incorporated in the amplitude formula. The temporal integrals can be expressed with the solutions of the free scalar field (\ref{eq:sol}). After we pass to the new variable of integration $s=e^{-\omega t}/\omega$ and calculate the bilateral derivative, the temporal integrals become:
\begin{equation}
\int_{0}^{\infty}ds s^{2}\left[-p\,'H_{-ik}^{(2)}(ps)\frac{\partial H_{-ik}^{(2)}(p\,'s)}{\partial (p\,'s)}+ p\, H_{-ik}^{(2)}(p\,'s)\frac{\partial H_{-ik}^{(2)}(ps)}{\partial (ps)}\right].
\end{equation}
Using the relation for derivative of Hankel functions (\ref{hn}), and the relation between Hankel functions and Bessel $K$ functions (\ref{hk}) from Appendix, we finally obtain the expression of transition amplitude in the form:
\begin{eqnarray}
&&\langle(\vec{p}\,)_{-},(\vec{p\,'})_{+}|\mathbf{S}_{1}|0\,\rangle= \frac{Ze^2i\omega}{4\pi^3|\vec{p}+\vec{p\,'}|^2}\nonumber\\
&&\times\biggl\{-p\,'\int_{0}^{\infty}ds s^2 \left[K_{-ik}(ips)K_{-ik-1}(ip\,'s)+K_{-ik}(ips)K_{-ik+1}(ip\,'s) \right] \nonumber\\
&&+p\int_{0}^{\infty}ds s^2 \left[K_{-ik}(ip\,'s)K_{-ik-1}(ips)+K_{-ik}(ip\,'s)K_{-ik+1}(ips)\right]\biggl\}.
\end{eqnarray}
The results of these integrals are discussed in Appendix (\ref{a0}). The final result for the amplitude is expressed in terms of hypergeometric Gauss functions, gamma Euler functions and unit step functions as follows:
\begin{eqnarray}\label{al1}
\langle(\vec{p}\,)_{-},(\vec{p\,'})_{+}|\mathbf{S}_{1}|0\,\rangle&=& -\frac{Ze^{2}m}{8\pi^3|\vec{p}+\vec{p\,'}|^{2}} \left[ \frac{1}{p^2}\theta(p-p\,')\left(-h_{k}\left(\frac{p\,'}{p}\right)+g_{k}\left(\frac{p\,'}{p}\right)\right)\right.\nonumber\\
&&\left.+ \frac{1}{p\,'^2}\theta(p\,'-p)\left(-g_{k}\left(\frac{p}{p\,'}\right)+h_{k}\left(\frac{p}{p\,'}\right)\right)\right],\label{an}
\end{eqnarray}
where the newly introduced functions $g_{k}\left(\frac{p}{p\,'}\right),\,h_{k}\left(\frac{p}{p\,'}\right)$ have the following expressions (with the specification that $g_{k}\left(\frac{p\,'}{p}\right),\,h_{k}\left(\frac{p\,'}{p}\right)$ is obtained when $p\rightleftarrows p\,'$):
\begin{eqnarray}\label{eq:gn}
g_{k}\left(\frac{p}{p\,'}\right)&=&\left(\frac{m}{\omega}\right)^{-1}\left[ \left(\frac{p}{p\,'}\right)^{-ik}\Gamma(1-ik)\Gamma(2+ik)\,_{2}F_{1}\left(2,1-ik;3;1-\left(\frac{p}{p\,'}\right)^{2}\right)\right.\nonumber\\
&&\left.+\left(\frac{p}{p\,'}\right)
^{-ik}\Gamma(2-ik)\Gamma(1+ik)\,_{2}F_{1}\left(1,2-ik;3;1-\left(\frac{p}{p\,'}\right)^{2}\right)\right]\nonumber\\
&&=\left(\frac{m}{\omega}\right)^{-1}\left[\left(\frac{p}{p\,'}\right)^{-ik}\Gamma(3)\Gamma(1-ik)\Gamma(ik)\,_{2}F_{1}\left(2,1-ik;1-ik;\left(\frac{p}{p\,'}\right)^{2}\right)\right.\nonumber\\
&&\left.+\left(\frac{p}{p\,'}\right)^{ik}\Gamma(3)\Gamma(2+ik)\Gamma(-ik)_{2}F_{1}\left(1,2+ik;1+ik;\left(\frac{p}{p\,'}\right)^{2}\right)\right.\nonumber\\
&&\left. +\left(\frac{p}{p\,'}\right)^{-ik}\Gamma(3)\Gamma(2-ik)\Gamma(ik)_{2}F_{1}\left(1,2-ik;1-ik;\left(\frac{p}{p\,'}\right)^{2}\right)
\right.\nonumber\\
&&\left. +\left(\frac{p}{p\,'}\right)^{ik}\Gamma(3)\Gamma(1+ik)\Gamma(-ik)\,_{2}F_{1}\left(2,1+ik;1+ik;\left(\frac{p}{p\,'}\right)^{2}\right)\right]
\end{eqnarray}
\begin{eqnarray}\label{eq:hn}
h_{k}\left(\frac{p}{p\,'}\right)&=&\left(\frac{m}{\omega}\right)^{-1}\left[\left(\frac{p}{p\,'}\right)^{-ik}\Gamma(1-ik)\Gamma(2+ik)\,_{2}F_{1}\left(1,1-ik;3;1-\left(\frac{p}{p\,'}\right)^{2}
\right)\right. \nonumber\\
&&\left.+\left(\frac{p}{p\,'}\right)^{-ik+2}\Gamma(2-ik)\Gamma(1+ik)\,{_2}F_{1}\left(2,2-ik;3;1-\left(\frac{p}{p\,'}\right)^{2}\right)\right]\nonumber\\
&&=\left(\frac{m}{\omega}\right)^{-1}\left[\left(\frac{p}{p\,'}\right)^{-ik}\Gamma(3)\Gamma(1-ik)\Gamma(1+ik)\,_{2}F_{1}\left(1,1-ik;-ik;\left(\frac{p}{p\,'}\right)^{2}\right)\right.\nonumber\\
&&\left.+\left(\frac{p}{p\,'}\right)^{2+ik}
\Gamma(3)\Gamma(2+ik)\Gamma(-1-ik)\,_{2}F_{1}\left(2,2+ik;2+ik;\left(\frac{p}{p\,'}\right)^{2}\right)\right.\nonumber\\
&&\left.+\left(\frac{p}{p\,'}\right)^{-ik+2}\Gamma(3)\Gamma(2-ik)\Gamma(-1+ik)\,_{2}F_{1}\left(2,2-ik;2-ik;\left(\frac{p}{p\,'}\right)^{2}\right)\right.\nonumber\\
&&\left.+\left(\frac{p}{p\,'}\right)^{ik}\Gamma(3)\Gamma(1-ik)\Gamma(1+ik)\,_{2}F_{1}\left(1,1+ik;ik;\left(\frac{p}{p\,'}\right)^{2}\right)\right]
\end{eqnarray}
The result of the integrals in the case $p\,'=p$, give terms proportional with delta Dirac function $\delta(p-p\,')$. However when we sum all the contributions of each integral with Hankel functions these terms cancel between them.

From equations (\ref{eq:gn}) and (\ref{eq:hn}), we can observe that the momentum is not conserved in the process of scalar pair production in external Coulomb field. The breaking of momentum conservation law is due to the external Coulomb field, because the geometry (\ref{metr}) has spatial translation invariance as an exact symmetry. If one had just the de Sitter background, without the Coulomb field, there would still be certain types of particle production as we will show in the next section, but no violation of momentum conservation law \cite{cd}.

It is clear that the contribution specific to the de Sitter geometry in our amplitude is contained in the functions $g_{k},\,h_{k}$. These quantities depend of the parameter $u=m/\omega$, and contain the effect of space expansion on the pair production process.
These functions will define the probability of pair production, which is the outcome of our perturbative approach.
The probability for pair production in Coulomb field is obtained by squaring the amplitude:
\begin{eqnarray}
&&\mathcal{P}_{i\rightarrow f}=|\langle(\vec{p}\,)_{-},(\vec{p\,'})_{+}|\mathbf{S}_{1}|0\,\rangle|^2=\frac{Z^{2}e^{4}m^{2}}{64{\pi}^{6}|\vec{p}+\vec{p\,'}|^{4}}\nonumber\\
&&\times\biggl\{\frac{1}{p^{4}}\theta(p-p\,')\left[\left|g_{k}\left(\frac{p\,'}{p}\right)\right|^{2}+\left|h_{k}
\left(\frac{p\,'}{p}\right)\right|^{2}-g_{k}\left(\frac{p\,'}{p}\right)h_{k}^{*}\left(\frac{p\,'}{p}\right)-h_{k}\left(\frac{p\,'}{p}\right)
g_{k}^{*}\left(\frac{p\,'}{p}\right)\right]\\ \nonumber
&&+\frac{1}{p\,'^4}\theta(p\,'-p)\left[\left|h_{k}\left(\frac{p}{p\,'}\right)\right|^{2}+\left|g_{k}\left(\frac{p}{p\,'}\right)\right|^{2}
-g_{k}\left(\frac{p}{p\,'}\right)h_{k}^{*}\left(\frac{p}{p\,'}\right)-h_{k}\left(\frac{p}{p\,'}\right)
g_{k}^{*}\left(\frac{p}{p^{'}}\right)\right]\biggl\}.\label{prob}
\end{eqnarray}
The total probability is obtained by integrating after the final momenta $p,\,p\,'$ the probability given above. This is a complicated task and we restrict to study the properties of the probability given in (\ref{prob}).
Since our calculations are exact we can study the results for probability in terms of the ratio mass of the particle/expansion factor.

\subsection{Graphical results}
\begin{figure}[h!t]
\includegraphics[scale=0.5]{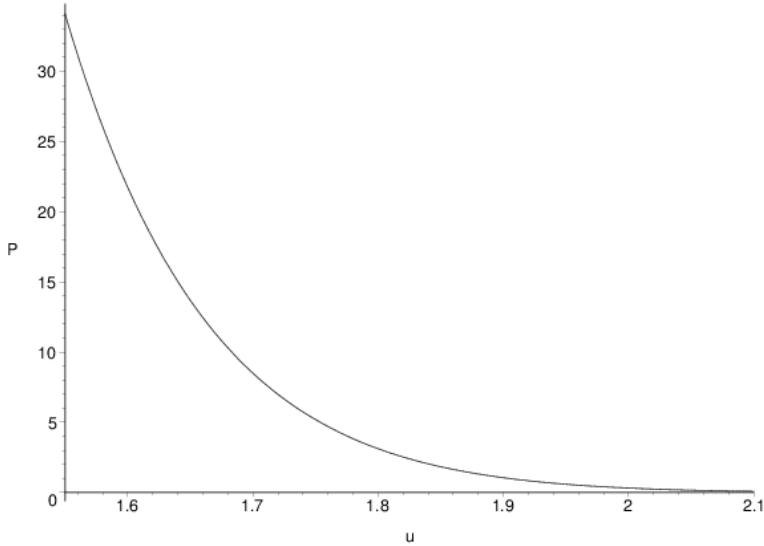}
\caption{$\mathcal{P}$ as a function of $u$ for $p/p\,'=0.1$.}
\label{f1}
\end{figure}

\begin{figure}[h!t]
\includegraphics[scale=0.5]{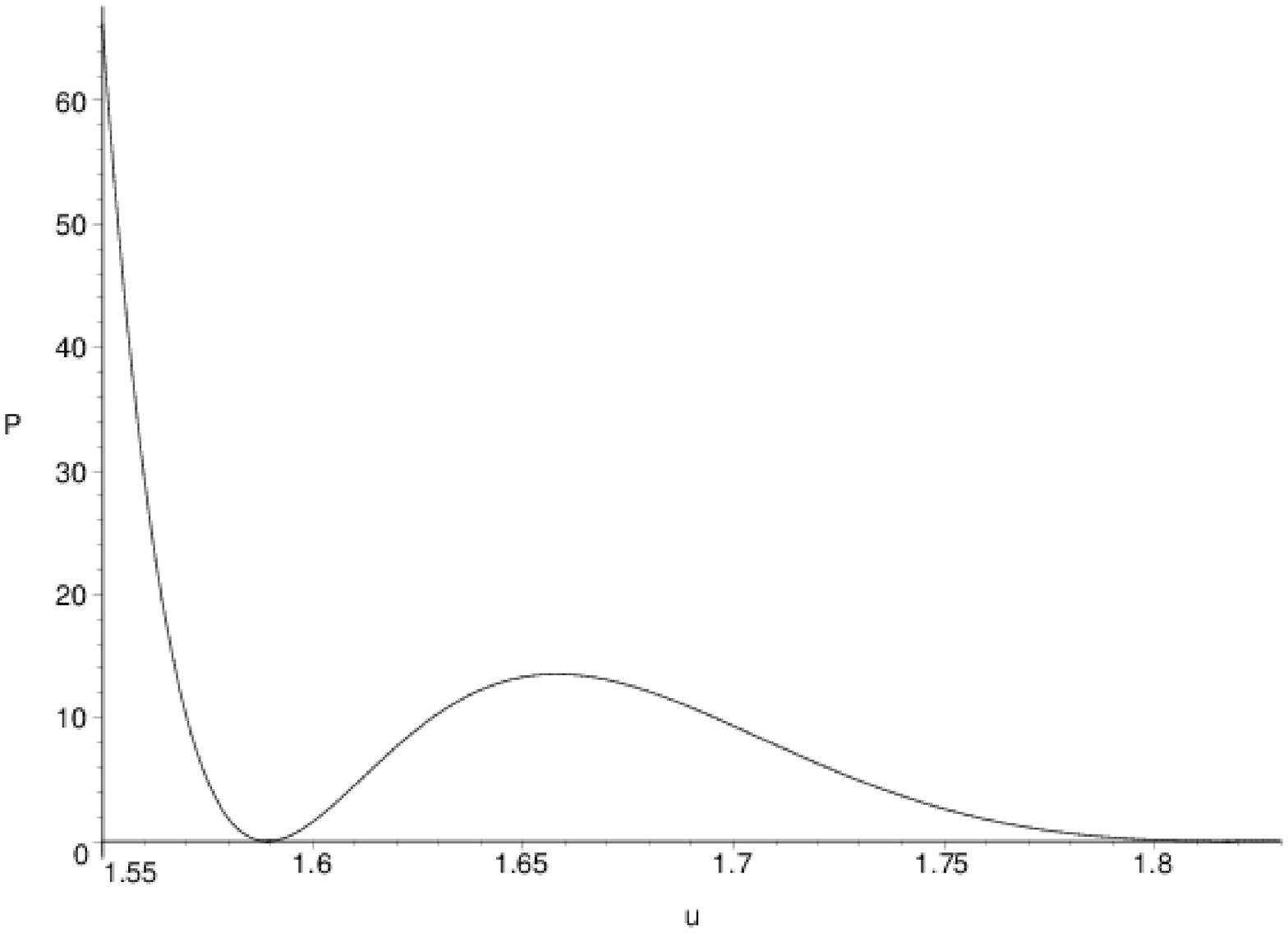}
\caption{$\mathcal{P}$ as a function of $u$ for $p/p\,'=0.001$.}
\label{f2}
\end{figure}

\begin{figure}[h!t]
\includegraphics[scale=0.5]{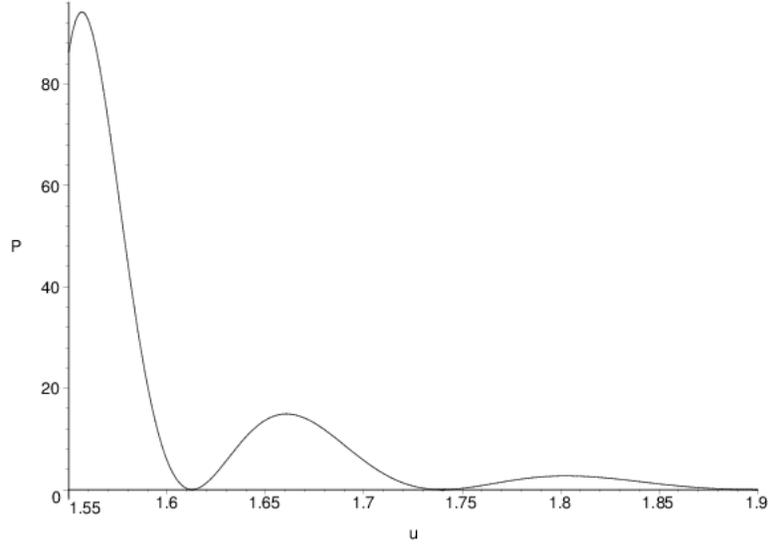}
\caption{$\mathcal{P}$ as a function of $u$ for $p/p\,'=0.00001$.}
\label{f3}
\end{figure}

In what follows we plot the probability as function of the parameter $u=m/\omega$ for different values of the momenta ratio $p\,'/p$. These results are presented in the Figs.(\ref{f1}-\ref{f3}).

Our results from the above graphs shows the dependence of the probability in terms of parameter $u=m/\omega$, for different values of the ratio $p/p\,'$. This result was obtained with the exact solution of Klein-Gordon equation expressed with Hankel function of imaginary index $H_{ik}^{(1)}(pz)$, where $k=\sqrt{(m/\omega)^2-9/4}$, solutions that are valid for $m/\omega>3/2$. In the above graphs the parameter $u$ was taken in the interval $u\in(3/2,\infty)$. Only for these values our analytical and graphical results are valid. We observe that the probabilities have a variation in terms of the ratio $p/p\,'$. For the momenta ratio closer to one the probability is very small as shown in Fig.(\ref{f1}). As the momenta ratio become very small the probability increase. This result show that the transitions from vacuum in which the momenta ratio are very small will be favoured in external Coulomb field (see Fig.(\ref{f3})). This is equivalent to say that there are bigger probabilities to produce pairs in which the momenta of the particle $p$, is very small comparatively with the momenta of the antiparticle $p\,'$  $(p\,'>> p)$.

Another observation that emerges from graphs Figs.(\ref{f1}-\ref{f3}) is that the probability decreases as the ratio $m/\omega$ increase. In other words, as the expansion parameter become smaller the probabilities for pair production reduces to zero. When the parameter $u=\infty$, which corresponds to the flat space limit (zero expansion factor), the probabilities are zero as we can observe from Figs.(\ref{f1}-\ref{f3}). This result is also obtained when we replace the ratio $m/\omega=\infty$ in the expressions for amplitudes and probabilities. We can draw the conclusion that in the Minkowski limit there are no production of scalar particles in external Coulomb field.

Further we can address the problem of separation between particles and antiparticles. The probability is proportional with the factor
$\frac{1}{|\vec{p}+\vec{p^{'}}|^{4}}$. The modulus from nominator can be expanded as:
\begin{equation}\label{sm}
|\vec{p}+\vec{p^{'}}|^{4}=(p^2+p\,'^2+2pp\,'\cos\beta)^2,
\end{equation}
where $\beta$ is the angle between the two momenta vectors. From here we can see that for $\beta=0$ the momenta of particle and antiparticle are parallel and have the same orientation. In this case the probability is minim and the pair will annihilate in vacuum. The opposite case is when $\beta=\pi$, which corresponds to the situation when the momenta of the particle and antiparticle are parallel and have opposite orientation. This is the case when the probability is maxim. Only in this case the particles and antiparticles could separate.

\section{Electromagnetic particle production}
The process: $vacuum\rightarrow \varphi+\varphi^*+\gamma$, in which scalar pair and a photon is generated from de Sitter vacuum is the topic of this section. This process is forbidden in scalar QED from flat space by the momentum and energy conservation laws. In de Sitter case the energy is no longer conserved and the probability for production of the triplet is no longer zero.
\subsection{Transition probability}
For computing the transition amplitude we will use the solution of the Klein-Gordon equation and the solution of Maxwell equation on de Sitter background, written in the momentum basis. The theory of Maxwell field in de Sitter geometry is simple is we recall the conformal invariance $A^{\mu}=\Omega^{-1}A^{\mu}_{M}$ (where $A^{\mu}_{M}$ is the vector potential in flat space). Then the fundamental solution of Maxwell equation in de Sitter geometry in momentum basis is \cite{Max}:
\begin{equation}
w^{i}_{\vec{q\,},\lambda}(t,\vec{x}\,)=e^{-2\omega\,t}
\frac{1}{(2\pi)^{3/2}}\frac{1}{\sqrt{2q}}\,e^{-iqt_{c}+i{\vec
q}\cdot {\vec x}}\,\varepsilon^{i}_{\lambda} (\vec{q\,})\,,
\end{equation}

The transition amplitude for this process reads:
\begin{equation}\label{tra}
\langle(\vec{p}\,)_{-},(\vec{p\,'})_{+},(\vec{q\,},\lambda)|\mathbf{S}_{1}|0\,\rangle=-e
\int\sqrt{-g(x)}\left[f^*_{\vec{p}^{\,\,\prime}}(x)\stackrel{\leftrightarrow}{\partial_{i}}f^*_{\vec{p} }\,(x)\right]w^i(x)^*d^{4}x.
\end{equation}
Our calculations will be valid only in the case $m/\omega> 3/2$ and we will use the solutions of the Klein-Gordon equation expressed in terms of Hankel functions of imaginary index.
Replacing all the quantities of interest in our amplitude we obtain that the spatial integrals give the delta Dirac function $\delta^{3}(\vec{p}+\vec{p^{'}}+\vec{q}\,)$, which assures the momentum conservation in this process. In the case of the temporal integrals we pass from Hankel functions to Bessel $J$ functions, arriving in this way at integrals (\ref{in1}-\ref{in2}), which are discussed in Appendix. The transition amplitude can be written passing to the new variable of integration $s=e^{-\omega t}/\omega$:
\begin{eqnarray}
&&\langle(\vec{p\,'}),(\vec{p}\,),(\vec{q},\lambda)|\mathbf{S}_{1}|0\,\rangle= -\frac{ie\pi(\vec{p}-\vec{p^{'}})\vec{\epsilon}\,\,^{*}_{\lambda}(\vec{q}\,)} {4\sqrt{2q}\,(2\pi)^{3/2}\sinh^{2}(\pi k)}\,\delta^{3}(\vec{p}+\vec{p^{'}}+\vec{q}\,)\int_{0}^{\infty}ds s\,e^{-iqs}\nonumber\\
&&\times[e^{\pi k}J_{-ik}(ps)J_{-ik}(p\,'s)+e^{-\pi k}J_{ik}(ps)J_{ik}(p\,'s)-J_{-ik}(ps)J_{ik}(p\,'s)-J_{ik}(ps)J_{-ik}(p\,'s)].
\nonumber\\
\end{eqnarray}
The final result is expressed in terms of Gauss hypergeometric functions $_{2}F_{1}$ and Appell hypergeometric functions of double argument $F_{4}$:
\begin{eqnarray}\label{av1}
\langle(\vec{p\,'}),(\vec{p}\,),(\vec{q},\lambda)|\mathbf{S}_{1}|0\,\rangle= -\frac{ie\pi}{4\sqrt{2}(2\pi)^{3/2}}(\vec{p}-\vec{p\,'})\vec{\epsilon^{*}_{\lambda}}(\vec{q}\,)\delta^{3}(\vec{p}+\vec{p\,'}+\vec{q})\nonumber\\
\times[h_{k}(p,p\,',q)
+h_{-k}(p,p\,',q)+f_{k}(p,p\,',q)+f_{-k}(p,p\,',q)].
\end{eqnarray}
The functions which define our amplitude are defined as follows:
\begin{eqnarray}\label{fv}
f_{k}(p,p\,',q)&=&\left(\frac{p}{p\,'}\right)^{ik}\frac{1}{\pi k \sinh(\pi k)q^{5/2}}F_{4}\left(1,\frac{3}{2};1+ik,1-ik;\frac{p^{2}}{q^2}+i0,\frac{p\,'^{2}}{{q}^{2}}+i0\right)\\ \nonumber &&+\left(\frac{p\,'}{p}\right)^{-ik}\frac{1}{\pi k \sinh(\pi k)q^{5/2}}F_{4}\left(1,\frac{3}{2};1-ik,1
+ik;\frac{p\,'^{2}}{q^2}+i0,\frac{p^{2}}{{q}^{2}}+i0\right)\label{f}
\end{eqnarray}

\begin{eqnarray}\label{hv}
h_{k}(p,p\,',q)&=& -\frac{ie^{-\pi k}q^{1/2}(k^2+1/4)}{4(pp\,')^{3/2}\sinh^{2}(\pi k)\cosh(\pi k)}\left[-ie^{\pi k}\,_{2}F_{1}\left(\frac{3}{2}+ik,\frac{3}{2}-ik;2;\frac{1-z}{2}\right) \right.\nonumber\\
&&\left.+_{2}F_{1}\left(\frac{3}{2}+ik,\frac{3}{2}-ik;2;\frac{1+z}{2}\right)\right]\label{h}
\end{eqnarray}
In equation (\ref{fv},\ref{hv}), the newly introduced notation $z$ is defined as:
\begin{equation}
z=\frac{p^2+p\,'^2-q^2}{2pp\,'}.
\end{equation}
We observe that the momentum is conserved in the process $vacuum\rightarrow \varphi+\varphi^*+\gamma$ . In the previous section we obtained that the process of pair production in external field broke the momentum conservation law. From here we can draw the conclusion that the particle production processes in the presence of gravity, preserve the momentum conservation law as long as there are no external fields (electric fields or magnetic fields).

The probability is obtained squaring the amplitude and summing after the photon polarizations $\lambda$. Since our amplitude is proportional with a delta Dirac term $\delta^3(\vec{p}\,)$ , by squaring this term we obtain $|\delta^3(\vec{p}\,)|^2=V\delta^3(\vec{p}\,)$. In this way we will define the probability per unit of volume:
\begin{eqnarray}\label{prv}
\mathcal{P}_{i\rightarrow f}&=&\frac{1}{2}\sum_{\lambda}\frac{|\langle(\vec{p\,'}),(\vec{p}\,),(\vec{q},\lambda)|\mathbf{S}_{1}|0\,\rangle|^2}{V}\nonumber\\ &=&\frac{1}{2}\sum_{\lambda}\frac{e^{2}[(\vec{p}-\vec{p\,'})\vec{\epsilon^{*}_{\lambda}}(\vec{q}\,)]^{2}\,\delta^{3}(\vec{p}+\vec{p\,'}+\vec{q}\,)}{64 \,(2\pi)^{3}}
\left[|h_{k}(p,p\,',q)|^{2}+ |h_{-k}(p,p\,',q)|^{2} \right.\nonumber\\
&&\left.+|f_{k}(p,p\,',q)|^{2}+ |f_{-k}(p,p\,',q)|^{2}
+h_{k}(p,p\,',q)h_{-k}^{*}(p,p\,',q)+h_{k}(p,p\,',q)f_{k}^{*}(p,p\,',q)\right.\nonumber\\
&&\left.+h_{k}(p,p\,',q)f_{-k}^{*}(p,p\,',q)
+h_{-k}(p,p\,',q)h_{k}^{*}(p,p\,',q)
+h_{-k}(p,p\,',q)f_{k}^{*}(p,p\,',q)\right.\nonumber\\
&&\left.+h_{-k}(p,p\,',q)f_{-k}^{*}(p,p\,',q)
+f_{k}(p,p\,',q)h_{k}^{*}(p,p\,',q)+f_{k}(p,p\,',q)h_{-k}^{*}(p,p\,',q)\right.\nonumber\\
&&\left.+f_{k}(p,p\,',q)f_{-k}^{*}(p,p\,',q)
+f_{-k}(p,p\,',q)h_{k}^{*}(p,p\,',q)+f_{-k}(p,p\,',q)h_{-k}^{*}(p,p\,',q)\right.\nonumber\\
&&\left.+f_{-k}(p,p\,',q)f_{k}^{*}(p,p\,',q)\right].
\end{eqnarray}

The probability depend on the factor: $\sum_{\lambda}|(\vec{p\,'}-\vec{p}\,)\vec{\epsilon^{*}_{\lambda}}(\vec{q\,})|^2$.
Using that the polarization vectors are orthogonal on photon momentum $\vec{q}\cdot\vec{\epsilon}_{\lambda}(\vec{q\,})=0$ and for any polarization $\lambda=\pm1$, they satisfy:
\begin{equation}
\sum_{\lambda}\varepsilon_{\lambda}({\vec q}\,)_i\,\varepsilon_{\lambda}({\vec
q}\,)^*_j=\delta_{ij}-\frac{q^i q^j}{q^2}\,,
\end{equation}
we can express the sum from probability taking into account that the momentum conservation as given by the delta Dirac term gives $\vec{p}=-(\vec{p\,'}+\vec{q}\,)$ :
\begin{eqnarray}
\sum_{\lambda}|(\vec{p\,'}-\vec{p}\,)\vec{\epsilon^{*}_{\lambda}}(\vec{q}\,)|^2=\sum_{\lambda}|2\vec{p\,'}\vec{\epsilon^{*}_{\lambda}}(\vec{q}\,)|^2
=4\left(\vec{p\,'}^2-\frac{(\vec{p\,'}\cdot\vec{q}\,)^2}{q^2}\right)=4p\,'^2\sin^2\theta_{qp\,'},
\end{eqnarray}
where $\theta_{qp\,'}$ is the angle between the momenta vectors $\vec{p\,'}$ and $\vec{q}$. This result shows that the probability will increase when the angle between the two vectors approach the value $\pi/2$ and vanishes for $\theta_{qp\,'}=0,\pi$. The conclusion is that the scalar particle with momentum $\vec{p}\,'$ and the photon will be emitted on different directions. The same conclusion emerges if we use the angle between $\vec{p}$ and $\vec{q}$.

The Minkowski limit can be obtained when the expansion factor vanishes or equivalently when $u=m/\omega=\infty$. In this limit the probability for generation of the triplet vanishes, thus obtaining the well known result from flat space where this process is forbidden by the energy conservation law.

\subsection{Graphical results}
In this subsection the probability given in Eq.(\ref{prv}) will be analysed in terms of the parameter $u=m/\omega$, using numerical values for the momenta $p\,,p\,'\,,q$ .
\begin{figure}[h!t]
\includegraphics[scale=0.5]{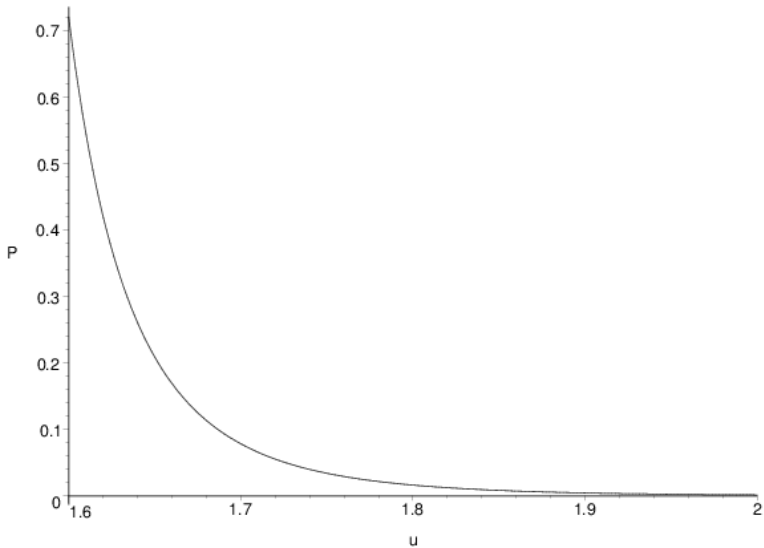}
\caption{$\mathcal{P}$ as a function of $u$ for $p=2,\,p\,'=2,\,q=1$.}
\label{f4}
\end{figure}

\begin{figure}[h!t]
\includegraphics[scale=0.5]{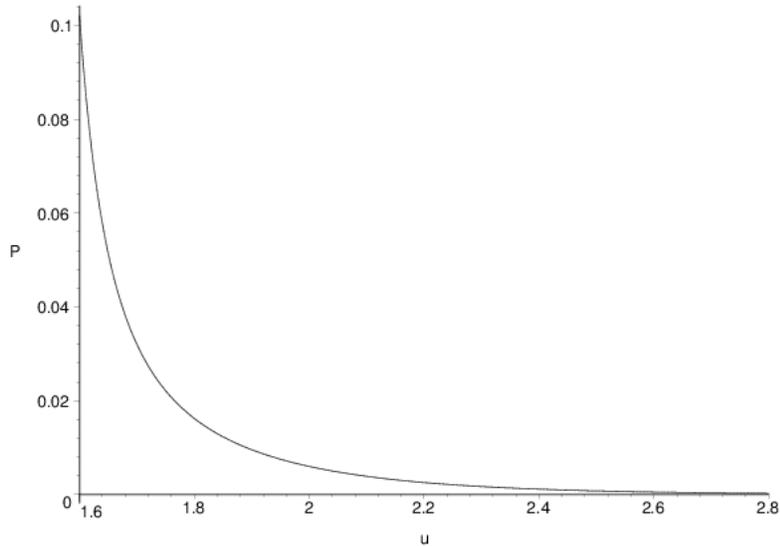}
\caption{$\mathcal{P}$ as a function of $u$ for $p=1,\,p\,'=2,\,q=2$.}
\label{f5}
\end{figure}

For plotting the probability we take the Appel hypergeometric functions $F_{4}$ to be equal with one, since these functions are less studied. Using the definition of functions $F_{4}$, one can observe that for large $u$ these functions are very convergent, and from here our approximation is well justified. Also one can observe from (\ref{fv}), that the shape of our graphs will be mainly determined by the factor $(k \sinh(\pi k))^{-1}$, which is very convergent at large $u$.
We observe from our graphs that the probability drops rapidly to zero when the expansion factor $\omega$ decreases. From graphs Figs.(\ref{f4}-\ref{f5}), it is clear that in the Minkowski limit the probability vanishes confirming in this way our analytical results for $u=\infty$.

\section{Conclusions}
We studied in this paper the interaction between Maxwell field and charged scalar field in de Sitter geometry using the perturbative methods as in flat space case. We established the equations of the interacting fields in the hypothesis of minimally coupling. Using then the solutions of the equations for interacting fields we construct the reduction formalism for the scalar field in de Sitter geometry. The definition of transition amplitudes was also obtained using the same perturbative methods as in flat space case. We compare two kinds of particle production in de Sitter geometry. Our study show that the particle production processes in external fields, do not preserve the momentum conservation law. Contrary to this we obtain that the mechanism of production of particles from de Sitter vacuum always preserve the momentum conservation law. Our analytical and graphical results prove that the probability of particle production is significant only in strong gravitational fields and vanishes in the flat space limit $\omega=0$. Indeed a direct perturbative calculation in Minkowski scalar QED prove that the amplitudes corresponding to our processes are zero.

\section{Appendix}
The main steps leading to amplitude of pair production in Coulomb field (\ref{al1}), will be detailed in this section. First, we will use the relation for the derivative of Hankel functions \cite{m3}:
\begin{equation}\label{hn}
\frac{dH_{\mu}^{(2)}(z)}{dz}=H_{\mu-1}^{(2)}(z)-H_{\mu+1}^{(2)}(z).
\end{equation}
and then the connection between Hankel functions and Bessel $K$ functions \cite{m1,m3}:
\begin{equation}\label{hk}
H^{(1,2)}_{\nu}(z)=\mp \left(\frac{2i}{\pi}\right)e^{\mp
i\pi\nu/2}K_{\nu}(\mp iz),
\end{equation}
arriving in this way at integrals of the type \cite{m3}:
\begin{eqnarray}\label{a0}
&&\int_0^{\infty} dz
z^{-\lambda}K_{\mu}(az)K_{\nu}(bz)=\frac{2^{-2-\lambda}a^{-\nu+\lambda-1}b\,^{\nu}}{\Gamma(1-\lambda)}\Gamma\left(\frac{1-\lambda+\mu+\nu}{2}\right)\Gamma\left(\frac{1-\lambda-\mu+\nu}{2}\right)\nonumber\\
&&\times\Gamma\left(\frac{1-\lambda+\mu-\nu}{2}\right)\Gamma\left(\frac{1-\lambda-\mu-\nu}{2}\right)
\,_{2}F_{1}\left(\frac{1-\lambda+\mu+\nu}{2},\frac{1-\lambda-\mu+\nu}{2};1-\lambda;1-\frac{b^2}{a^2}\right),\nonumber\\
&&Re(a+b)>0\,,Re(\lambda)<1-|Re(\mu)|-|Re(\nu)|.
\end{eqnarray}
In our case $\lambda=-2$ and the second condition for convergence
is satisfied. We also observe that in our case $a,b$ are complex
and for solving our integrals we add to $a$ a small real part
$a\rightarrow a+\epsilon$, with $\epsilon>0$ and in the end we
take the limit $\epsilon\rightarrow 0$. This assure the
convergence of our integral and will correctly define the unit
step functions and $h_{k},\,g_{k}$ functions.

The final form of the functions $g_{k}\,,h_{k}$ which define the amplitude of pair production in Coulomb field is obtained if we use the following identity between hypergeometric functions \cite{m3,m1}:
\begin{eqnarray}\label{hy}
_{2}F_{1}(a,b;c;z)&=& \frac{\Gamma(c)\Gamma(c-a-b)}{\Gamma(c-a)\Gamma(c-b)}\,_{2}F_{1}(a,b;a+b-c+1;1-z)\\ \nonumber
&&+(1-z)^{c-a-b}\,\frac{\Gamma(c)\Gamma(a+b-c)}{\Gamma(a)\Gamma(b)}\,_{2}F_{1}(c-a,c-b;c-a-b+1;1-z).
\end{eqnarray}

For the second process of particle production from de Sitter vacuum, we use the well known formula for Hankel functions in terms of Bessel $J$ functions \cite{m1,m2,m3}:
\begin{eqnarray}
H^{(1)}_{\mu}(z)=\frac{J_{-\mu}(z)-e^{-i\pi\mu}J_{\mu}(z)}{i\sin(\pi\mu)}\nonumber\\
H^{(2)}_{\mu}(z)=\frac{e^{i\pi\mu}J_{\mu}(z)-J_{-\mu}(z)}{i\sin(\pi\mu)}.
\end{eqnarray}
together with:
\begin{equation}
K_{\frac{1}{2}}(z)=\sqrt{\frac{\pi}{2z}}e^{-z},
\end{equation}
and we arrive at the following integrals \cite{m3}:
\begin{equation}\label{in1}
\int_{0}^{\infty}dx x^{3/2}K_{\frac{1}{2}}(cx)J_{\nu}(p\,x)J_{\nu}(p\,'x)=-\frac{c^{1/2}}{\sqrt{2\pi}(pp\,')^{3/2}}\frac{d}{du}Q_{\nu-\frac{1}{2}}(u)
\end{equation}
which depends on variable $u=\frac{a^2+b^2+c^2}{2ab}$.

The second type of integrals used are \cite{m3}:
\begin{eqnarray}\label{in2}
\int_{0}^{\infty}x^{3/2}K_{\frac{1}{2}}(cx)J_{\nu}(ax)J_{-\nu}(bx)dx =\left(\frac{a}{b}\right)^{\nu}\frac{\sin\pi\nu}{\sqrt{2\pi}\nu}\\ \nonumber
\times F_{4}\left(1,\frac{3}{2};1+\nu,1-\nu;-\frac{a^{2}}{c^{2}},-\frac{b^{2}}{c^{2}}\right).
\end{eqnarray}
These integrals are convergent for $Re(c)>0$. For this reason we must replace $c\rightarrow\epsilon-iq\,,\epsilon>0$, when we evaluate these integrals. In the end we take the limit $\epsilon\rightarrow0$.

For establishing the final results we use the following relations for Legendre functions of second kind \cite{m3}:
\begin{eqnarray}
Q_{\nu}(z\pm i0)= \frac{\pi}{2\sin(\pi\nu)}\left[e^{\mp i\pi\nu}P_{\nu}(z)-P_{\nu}(-z)\right],
\end{eqnarray}
when $-1<z<1$ and their relations with hypergeometric functions $_{2}F_{1}$ \cite{m3}
\begin{equation}
\frac{d}{dz}Q_{\nu}(z\pm i0)=\frac{\pi\nu(\nu+1)}{4\sin(\pi\nu)}\left[e^{\mp i\pi\nu}\,_{2}F_{1}\left(1-\nu,2+\nu;2;\frac{1-z}{2}\right)+
_{2}F_{1}\left(1-\nu,2+\nu;2;\frac{1+z}{2}\right)\right].
\end{equation}
The Legendre functions of first kind can be written in terms of hypergeometric functions $_{2}F_{1}$ \cite{m3}:
\begin{equation}
 P_{\nu}(z)= \,_{2}F_{1}\left(-\nu,1+\nu;1;\frac{1-z}{2}\right).
\end{equation}

\textbf{Acknowledgements}

Mihaela-Andreea B\u aloi was supported by the strategic grant POSDRU/159/1.5/S/137750, Project "Doctoral and Postdoctoral programs support for increased competitiveness in Exact Sciences research" cofinanced by European Social Fund within the Sectoral Operational Programme Human Resources Development 2007-2013.

Cosmin Crucean was supported by a grant of the Romanian National Authority for Scientific Research,
Programme for research-Space Technology and Advanced Research-STAR, project nr. 72/29.11.2013 between
Romanian Space Agency and West University of Timisoara.

\end{document}